\documentclass{phostproc}

\usepackage{graphicx}
\usepackage{hyperref}
\usepackage[]{natbib}  
\usepackage{pgf, tikz}
\usepackage{amssymb}

\title{The effects of rotation on wave-induced transport in stars: the case of an intermediate-mass star}
\author{Q. Andr\'e$^{1},$ 
        S. Mathis$^{1}$
        and L. Amard$^{2}$}

\affiliation{$^{1}$ AIM, CEA, CNRS, Universit\'e Paris-Saclay, Universit\'e Paris Diderot, Sorbonne Paris Cit\'e, F-91191 Gif-sur-Yvette, France \\
			 $^{2}$ University of Exeter, Department of Physics \& Astronomy, Stoker Road, Devon, Exeter, EX4 4QL, UK}

\shorttitle{Effect of rotation on AM transport by waves}
\shortauthors{Q. Andr\'e, S. Mathis \& L. Amard}

\abs{
Internal gravity waves propagate in stellar radiative zones and transport angular momentum throughout the evolution of stars, shaping the internal rotation profile of these regions. We use the analytical study of \citet{AndreEtal2018_pnps} to assess the impact of rotation and differential rotation on the thermal dissipation of internal gravity waves near the equatorial plane, without making any assumption on the relative strength of rotation relative to that of the background stable stratification. {This dissipation is one of the mechanisms allowing a wave-induced deposition/extraction of angular momentum.} Here, we apply the analysis to a 3 solar mass star of which we have computed the one-dimensional evolution, from the pre-main sequence to the end of the main sequence. We show that rotation and differential rotation do not significantly modify the efficiency of the damping of internal gravity waves by thermal diffusion, except very close to their excitation region.
}

\begin{document}

\maketitle

\section{Introduction}

Intermediate-mass and massive stars have a convective core surrounded by a stably stratified, radiative envelope. Reynolds stresses in the core, and penetrative convection or overshoot at the core boundary, provide a mechanical forcing that can sustain the excitation of internal gravity waves \citep{LoSchatzman1997,BrowningEtal2004,SamadiEtal2010,NeinerEtal2012,RogersEtal2013,ShiodeEtal2013,AntociEtal2014,AertsRogers2015,AugustsonEtal2016,BowmanEtal2018}, which then propagate and redistribute angular momentum (AM hereafter) in the radiative envelope \citep[e.g.][]{LeeSaio1993,PantillonEtal2007,Rogers2015}. Waves that undergo constructive interferences lead to the creation of global modes of oscillation, that eventually {are} observed by asteroseismology. The properties of these oscillations help constraining the internal rotation profiles of the stars observed {using} asteroseismic methods \citep{AertsEtal2010}.

First, weak differential rotation rates have been found in the radiative envelope of intermediate-mass and massive stars \citep{KurtzEtal2014,SaioEtal2015,TrianaEtal2015,MurphyEtal2016,VanReethEtal2016,AertsEtal2017,AertsEtal2018,OuazzaniEtal2018}.
In addition, \cite{BeckEtal2012,MosserEtal2012,DeheuvelsEtal2012,DeheuvelsEtal2014,DeheuvelsEtal2015,SpadaEtal2016,GehanEtal2018} found, by measuring their core-to-surface rotation ratio, that low-mass subgiant and red giant stars {are also the seat of weak radial gradients of rotation}.
Finally, a strong extraction of AM is required to explain the rotation rates of white dwarfs \citep[e.g.][]{SuijsEtal2008,HermesEtal2017} and neutron stars \citep[e.g.][]{HegerEtal2005,HirschiMaeder2010}. These observational constraints indicate that stellar radiative zones are the seat of physical processes efficient at redistributing AM for a wide range of masses and all along their evolution.
Several theoretical studies have proposed that internal gravity waves could play an important role in extracting AM from the radiative zone in which they propagate \citep[e.g.][]{Schatzman1993,LeeSaio1993,ZahnEtal1997,TalonCharbonnel2005,FullerEtal2014,Rogers2015,PinconEtal2017}.

Because stars are possibly rapidly rotating, the Coriolis and centrifugal accelerations provide additional restoring forces to internal gravity waves, which then have to be treated as gravito-inertial waves \citep{DintransRieutord2000}. However, state-of-the-art analytical prescriptions implemented in one-dimensional (1D) stellar evolution codes have  generally been derived in the framework of pure gravity waves, thus neglecting the action of the Coriolis acceleration \citep[e.g.][]{ZahnEtal1997}.
{In addition, most of the works studying AM transport by gravito-inertial waves have studied cases for which the buoyancy force is strong compared to the Coriolis acceleration in the radial direction, using the Traditional Approximation of Rotation \citep[TAR, e.g.][]{PantillonEtal2007,Mathis2009,MathisEtal2018}}.
It is thus important to assess precisely the impact of rotation and differential rotation upon the excitation, propagation and dissipation of internal gravity waves in stellar radiative zones for any relative strength of the stratification \citep[e.g][]{MirouhEtal2016,Prat2018}, because this could affect the efficiency of AM exchanges at work, and the resulting rotation profiles of these regions. In this proceeding paper, we address the question of the influence of rotation and differential rotation upon the efficiency of thermal diffusion {that damp} internal gravity waves, {which is with corotation resonances and wave nonlinear breaking leading to wave-induced transport of AM.} 
We apply the analysis of \cite{AndreEtal2018_pnps} to the case of an intermediate-mass star.

\section{Physical formulation of the problem}
In a previous proceeding paper \citep{AndreEtal2018_pnps}, we have presented an analytical analysis aiming at assessing the effect of rotation (and differential rotation) on the dissipation of gravito-inertial waves by thermal diffusion. Following an early work by \cite{Ando1985}, the analysis was carried out near the equatorial plane of a star, {that also allows us to probe the effects of the non-traditional terms of the Coriolis acceleration \citep[e.g.][]{GerkemaEtal2008}}. Here, we recall the main steps, but we refer to \citet{AndreEtal2018_pnps} for more {mathematical} details. {First, we note that one main effect of the Coriolis acceleration {in rapidly rotating stars} is to confine the waves near the equator \citep[e.g.][]{LeeSaio1997}. Thus, relevant dynamical features are expected to be captured in such an equatorial model.} {It also allows us to filter, as a first step, more complex three-dimensional behaviours such as this latitudinal trapping in the presence of (differential) rotation \citep[e.g.][]{Prat2018}.} In addition, we note that such a (quasi) two-dimensional equatorial model has proven itself useful to reproduce asteroseismic constraints on intermediate-mass and massive stars' rotation profile with nonlinear numerical simulations of wave-mean flow interaction \citep{Rogers2015}. Therefore, an equivalent analytical model such as the one we use here could be helpful to interpret results from these two-dimensional simulations.

When including thermal diffusion – characterised by a thermal diffusivity factor $\kappa$ – in the heat transport equation, the AM flux, carried by the wave is decreased along its propagation. By making a quasi-adiabatic analysis of the primitive equations \citep[][]{Press1981,ZahnEtal1997,AndreEtal2018_pnps}, that is justifed in radiative regions with low Prandtl numbers, one can derive that this decrease follows an exponential low, with a characteristic damping factor $\tau$:

\begin{equation}
    \mathcal{F}_J(r) =  \mathcal{F}_J(r_c)\exp\left\{-\tau(r)\right\},
\end{equation}
where $\mathcal{F}_J$ is the AM flux carried by the wave, and $r_c$ is the radius of excitation of the wave {at the convective core/radiative envelope boundary}. The expression of the damping factor $\tau$ can in turn be written involving a characteristic length, $L$, such that
\begin{equation}
    \tau(r) = \int^r_{r_{c}} \frac{\textrm{d}r'}{L(r')},
    \label{eq:tau}
\end{equation}
as in \cite{FullerEtal2014}. This so-called {wave penetration length} provides a first proxy that we can focus on to examine the effects of rotation and differential rotation upon the wave-induced transport {of AM}, as in \citet{AndreEtal2018_pnps}. This represents an evaluation of a (local) characteristic length over which a wave will be able to exchange AM with the background flow. It thus provides a proxy that assesses the efficiency of the coupling between the waves' excitation region (convective cores) and stably stratified radiative regions in which they propagate.

Doing the algebra along these lines, we have shown in \citet[][]{AndreEtal2018_pnps} that the wave penetration length, normalised by that of pure gravity waves $L_0$, is given {near the equatorial plane} by
\begin{equation}
    \frac{L}{L_0} = \frac{\left(\displaystyle 1 + \alpha \frac{\textrm{S}(\textrm{S}+\textrm{Ri}^{-1/2})}{1-\textrm{Fr}^2}\right)^{1/2}}
        {\displaystyle 1 + \alpha \, \textrm{S}(\textrm{S}+\textrm{Ri}^{-1/2})},
      \label{eq:L}
\end{equation}
where
\begin{equation}
  \alpha = 1 - \left(\frac{m}{l}\right)^2,
  \label{eq:alpha}
\end{equation}
and the dimensionless Froude numbers and Richardson number {have been introduced with their} following expressions:
\begin{equation}
\textrm{Fr} = \frac{\omega}{N}, ~~~~~ \textrm{S} = \frac{2\Omega}{N}, ~~~~~ \textrm{and} ~~~~~ \textrm{Ri} = \left(\frac{N}{r\textrm{d}\Omega/\textrm{d}r}\right)^2.
\end{equation}
In Eq.~(\ref{eq:alpha}) above, $m$ is the azimuthal wave number and $l$ the angular degree of the wave.
{As already stated in \citet[][]{AndreEtal2018_pnps}, one can see that the parameter $\alpha$ somewhat weights the terms linked to rotation.  Waves with $l=m$ behave like pure gravity waves in the framework of our analysis, because then $\alpha = 0$ and thus $L=L_0$. Therefore, because waves with $m=0$ do not lead to a net transport of AM, we rather expect waves with $m=1$ and $l>1$ to be the more impacted by rotation.}

\section{Application to a 3 solar mass star}\label{sec3Msun}
\begin{figure*}[ht]
    \centering
    \includegraphics[height=6.35cm]{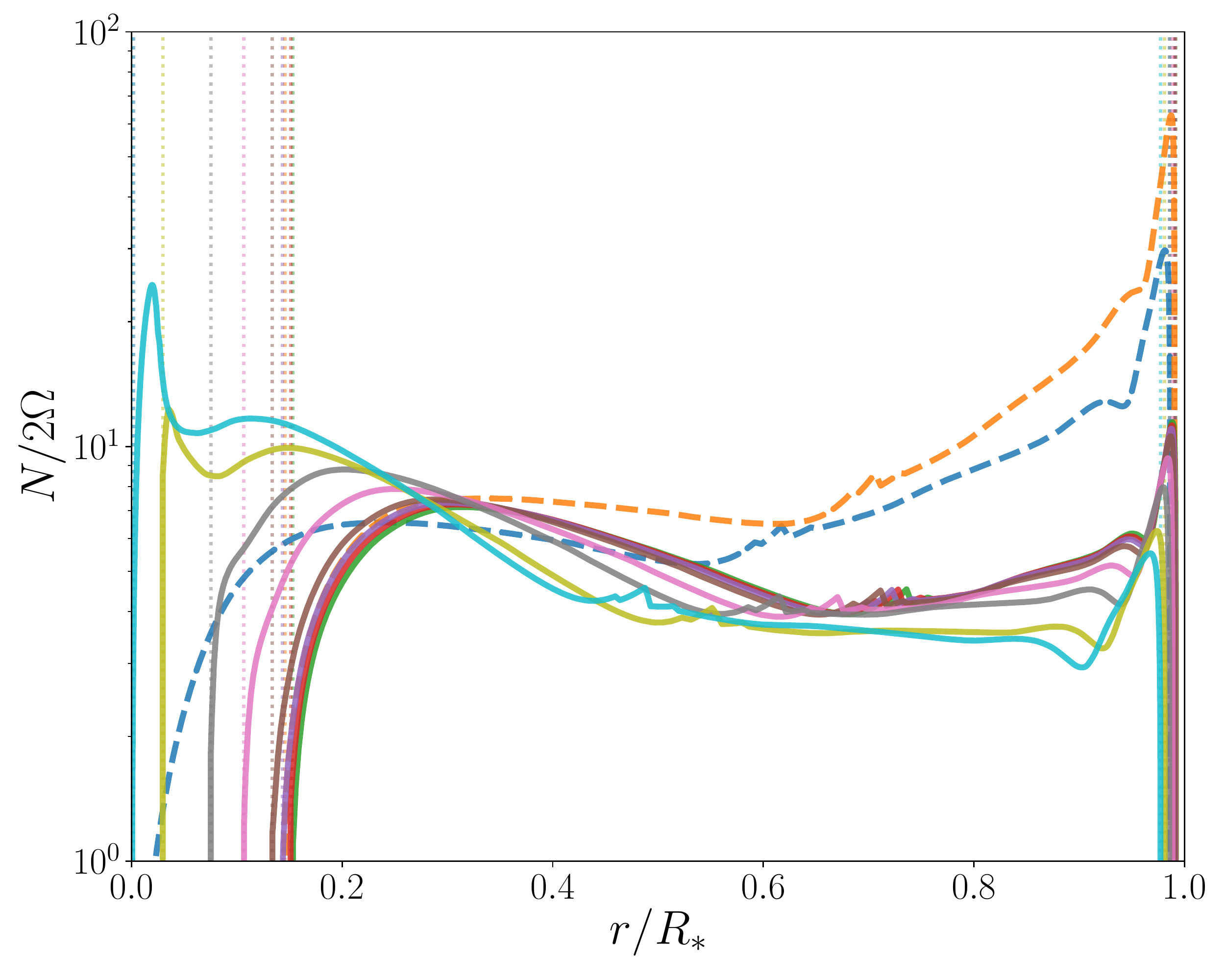}\hspace{-0.15cm}
    \includegraphics[height=6.35cm]{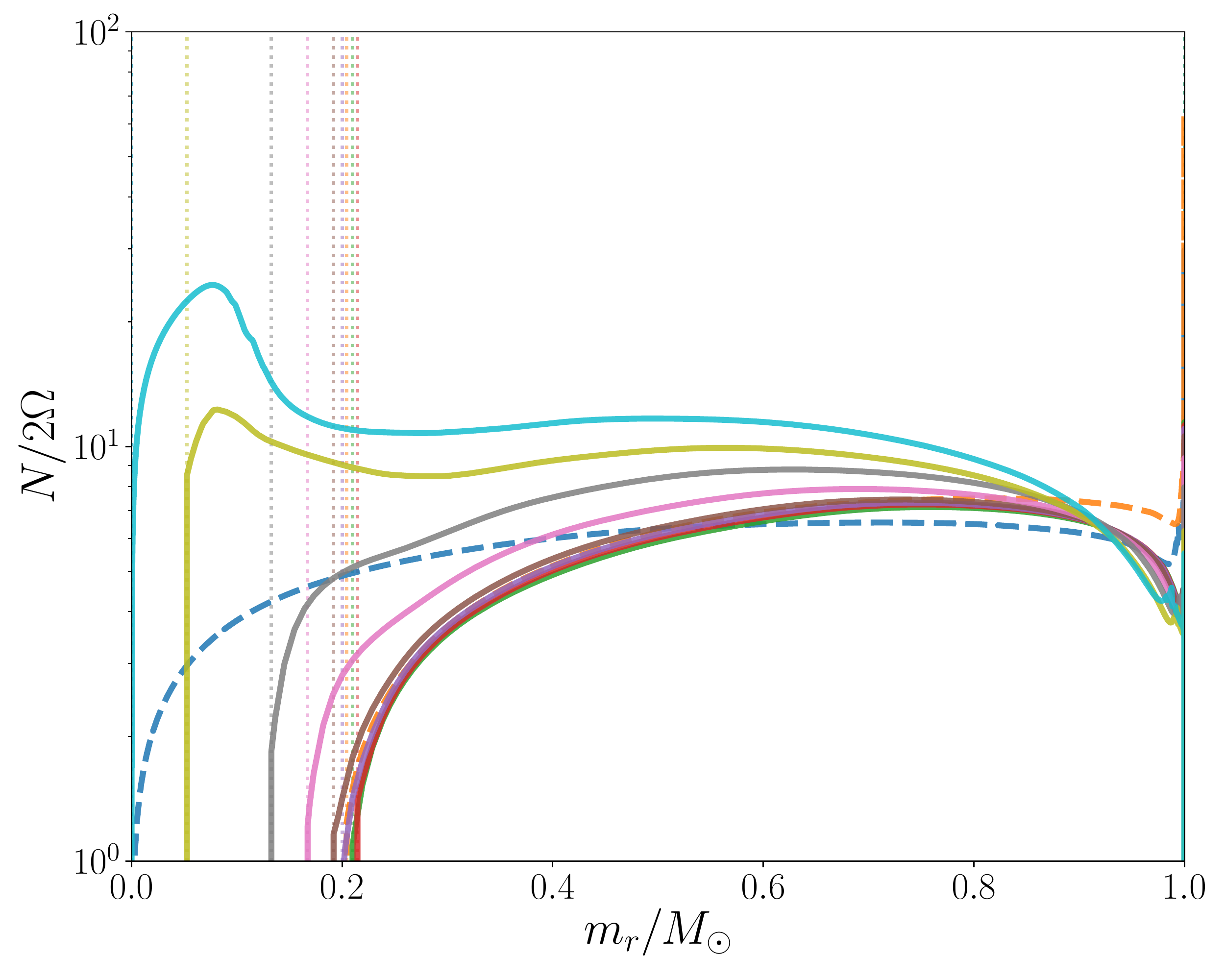}\hspace{-0.2cm}
    \begin{tikzpicture}[scale=1.0]
     \node[color=white] at (0,0) {o};
     \node[anchor=south west,inner sep=0] at (0,1.4) {\includegraphics[height=3.6cm]{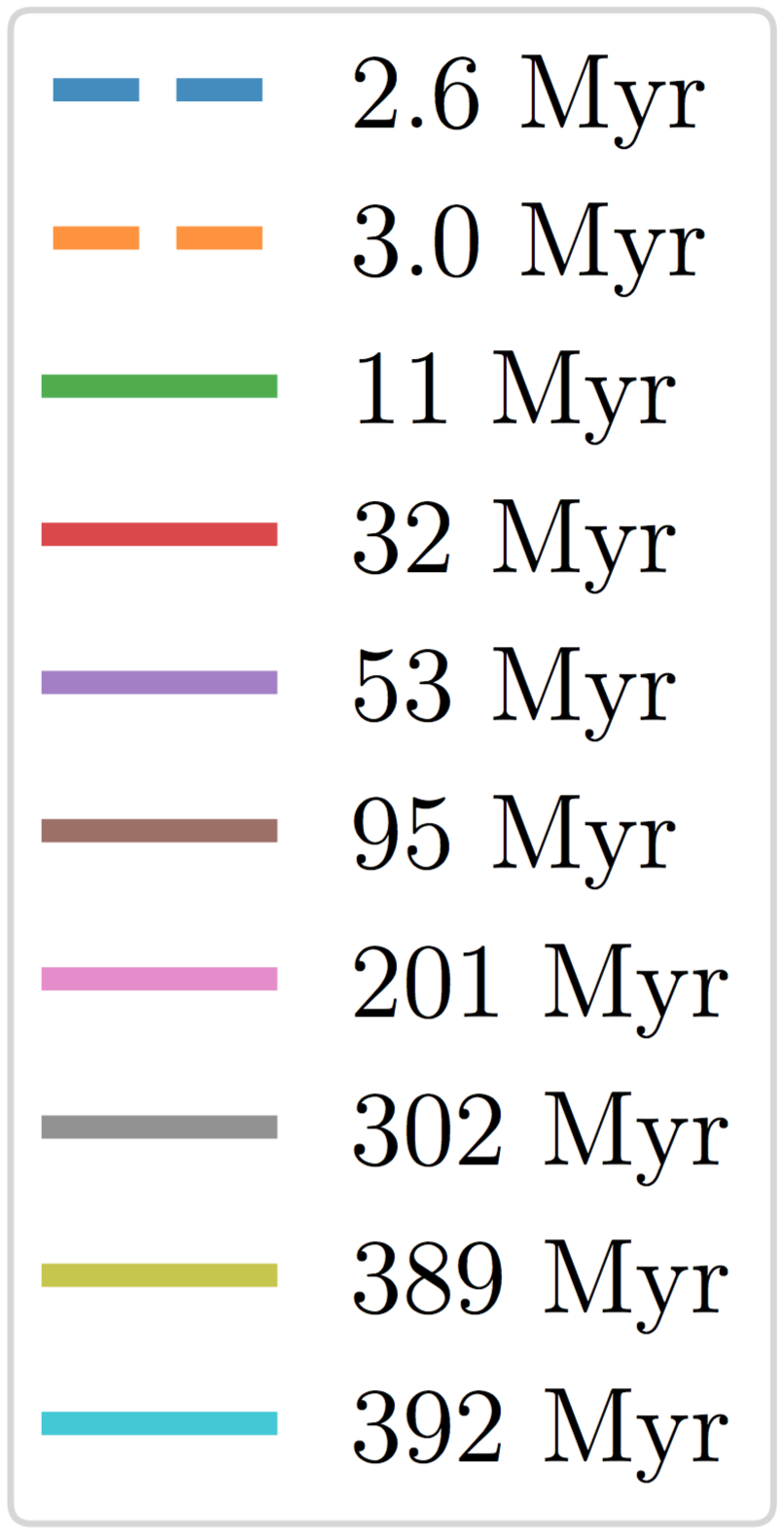}};
    \end{tikzpicture}\\[-0.3cm]
    \caption{Ratio $N/2\Omega$ as a function of normalised radius (\textit{left panel}) and normalised mass (\textit{right panel}). The different curves correspond to different ages in the {evolution}. Dashed lines correspond to the PMS (before the ZAMS at 4\,Myr), and solid lines to the MS, until the TAMS at 392\,Myr. Vertical dotted lines correspond to the boundaries of the stably stratified region, at corresponding ages.}
    \label{fig:S}
    \vspace{0.4cm}
    \centering
    \includegraphics[height=6.35cm]{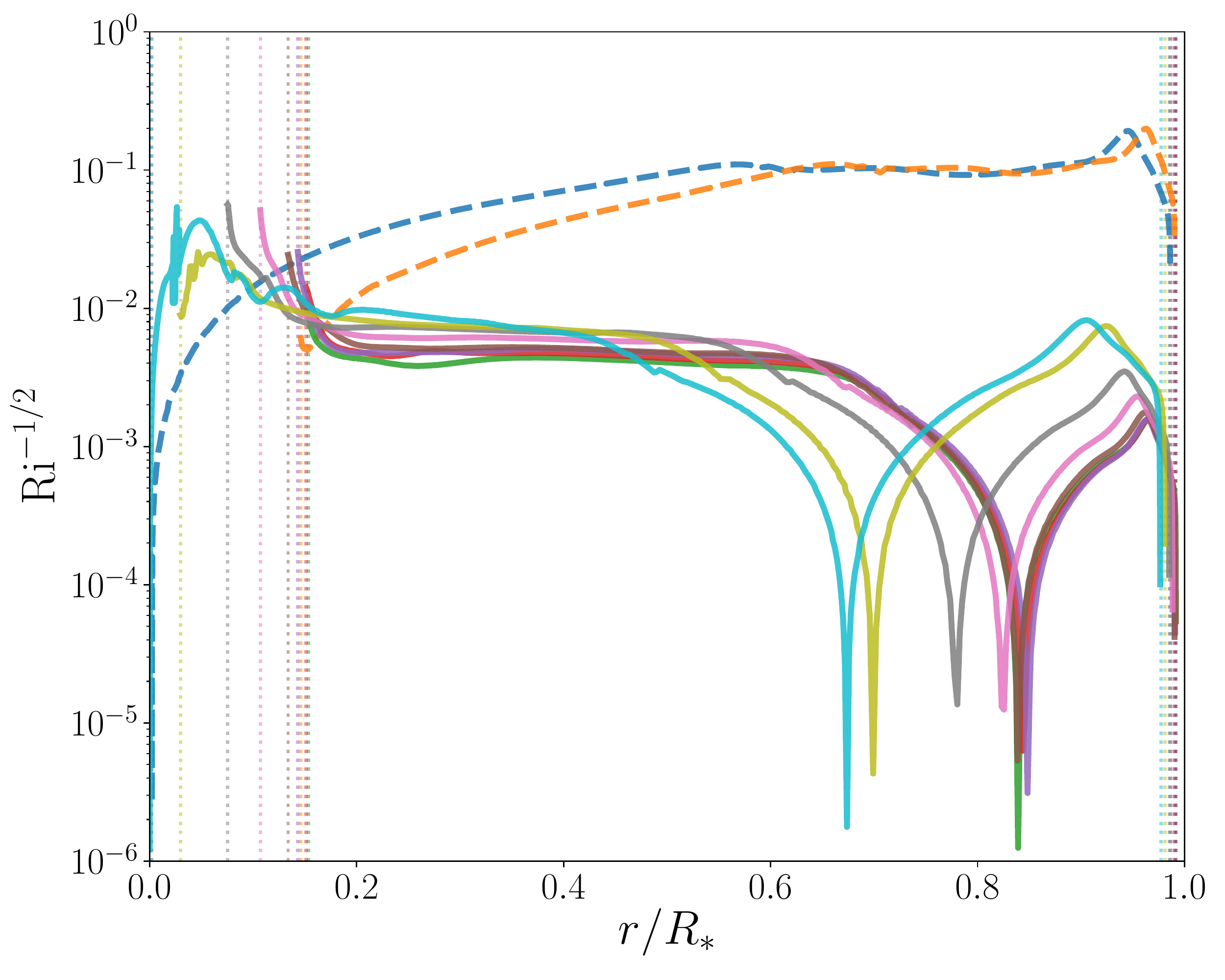}\hspace{-0.15cm}
    \includegraphics[height=6.35cm]{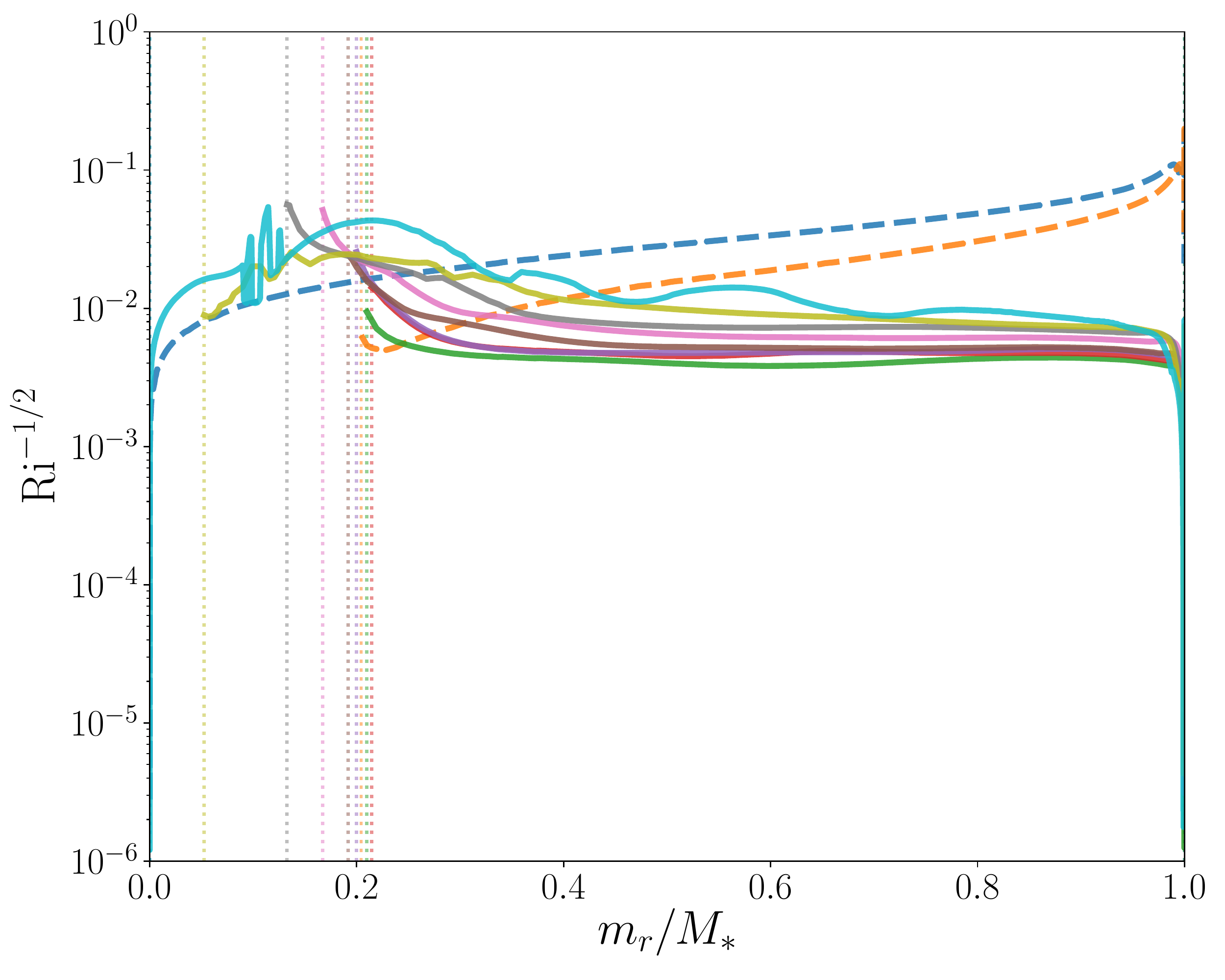}\hspace{-0.2cm}
    \begin{tikzpicture}[scale=1.0]
     \node[color=white] at (0,0) {o};
     \node[anchor=south west,inner sep=0] at (0,1.4) {\includegraphics[height=3.5cm]{colorbar.png}};
    \end{tikzpicture}\\[-0.3cm]
    \caption{Same as Fig.~\ref{fig:S} but for Ri$^{-1/2}$.}
    \label{fig:Rim12}
\end{figure*}

\subsection{Evolution models}
We now want to explore how the normalised wave penetration length, $L/L_0$, varies along the evolution of an intermediate-mass star. We have computed the one-dimensional evolution model of a $3\,M_{\odot}$ star at solar metallicity, using the stellar evolution code STAREVOL \citep[e.g.][Amard \textit{et al.}, in prep]{SiessEtal2000,PalaciosEtal2003,TalonCharbonnel2005,DecressinEtal2009,LagardeEtal2012,CharbonnelEtal2013,AmardEtal2016}. It computes AM evolution (internal transport and surface extraction) in a self-consistent way during stellar evolution. {We choose here to only take into account} AM transport due to meridional circulation and shear instabilities {\citep{Zahn1992,MaederZahn1998,MathisZahn2004}}. This allows us to isolate the needed effect of internal gravity waves (and potential other missing transport processes), since these mechanisms are not sufficient on their own to reproduce the observations \citep[e.g.][]{EggenbergerEtal2012,MarquesEtal2013,CeillierEtal2013,MathisEtal2018}. The initial rotation rate was taken to be 50\% of the surface critical velocity, defined by $v_{\scriptsize \textrm{crit}} = \sqrt{\mathcal{G}M_*/R_*}$. {This is slightly lower than the median value for $3\,M_{\odot}$ stars, which from \cite{ZorecRoyer2012} is around 65\% of the critical velocity.}

The radial profiles of the parameters relevant to our study, obtained from these models, are shown on Figs. \ref{fig:S} and \ref{fig:Rim12} for ages ranging from 2.6\,Myr to 392\,Myr (terminal age main sequence; TAMS hereafter), the zero age main sequence (ZAMS) occurring at 4.0\,Myr in our model. Those figures, and all subsequent figures, are shown as a function of radius normalised by stellar radius $R_*$ (left panels), and integrated mass $m_r$ normalised by stellar mass $M_*$ (right panels), at each specific age. Plotting quantities as a function of the integrated mass in addition of radius, allows to stretch the inner region, which contains the wave excitation region where rotation effects are expected to be more important \citep{AndreEtal2018_pnps}. Figure \ref{fig:S} displays the ratio $N/2\Omega = S^{-1}$, while Fig.~\ref{fig:Rim12} displays Ri$^{-1/2}$, as this is as it appears in Eq.~\ref{eq:L}. On those panels, dashed lines correspond to profiles on the PMS, and solid lines correspond to profiles on the MS. Finally, vertical dotted lines indicate the inner and outer boundaries of the radiative region. We note that some models present convective surface layers.

The ratio $N/2\Omega$ represents the relative magnitude of two characteristic frequencies: the buoyancy frequency, $N$, and the Coriolis frequency, $2\Omega$. When $N \gg 2\Omega$, rotation effects can be neglected and gravito-inertial waves behave like pure gravity waves. On Fig.~\ref{fig:S}, one can see that during the PMS, when the convective core grows, this ratio is close to $\sim 10$ on average in the bulk of the radiative zone, while on the MS it is close to $\sim 5$. On the MS, the profile in the bulk of the radiative zone stays fairly stable until 100 Myr of evolution, with a maximum value around 7. Thus, rotation effects are worth focusing on in the case of $3\,M_{\odot}$ star, like in the case of solar-type stars on the PMS \citep{AndreEtal2018_pnps}, since the ratio $N/2\Omega$ can be of order unity {during the main sequence, especially near the convective core, where the buoyancy frequency goes to zero}. For {more advanced evolutionary stages}, as the convective core shrinks, the radiative region propagates down leaving a higher buoyancy frequency in the inner region, due to chemical stratification.
Now, let us consider the behaviour of the Richardson number. When  $\textrm{Ri}^{-1/2} \gg 1$, differential rotation can potentially have a strong impact upon the dynamics and {thermal} diffusion of internal gravity waves. On Fig.~\ref{fig:Rim12}, one can see that during the PMS, this number reaches a value of $10^{-1}$ but is always lower to $10^{-1}$ during the rest of the evolution of the star. Thus, like in the solar-type stars' case \citep[][]{AndreEtal2018_pnps}, we do not expect this term to play a significant role.

\subsection{Results}
\begin{figure*}[ht]
    \centering
    \includegraphics[height=6.35cm]{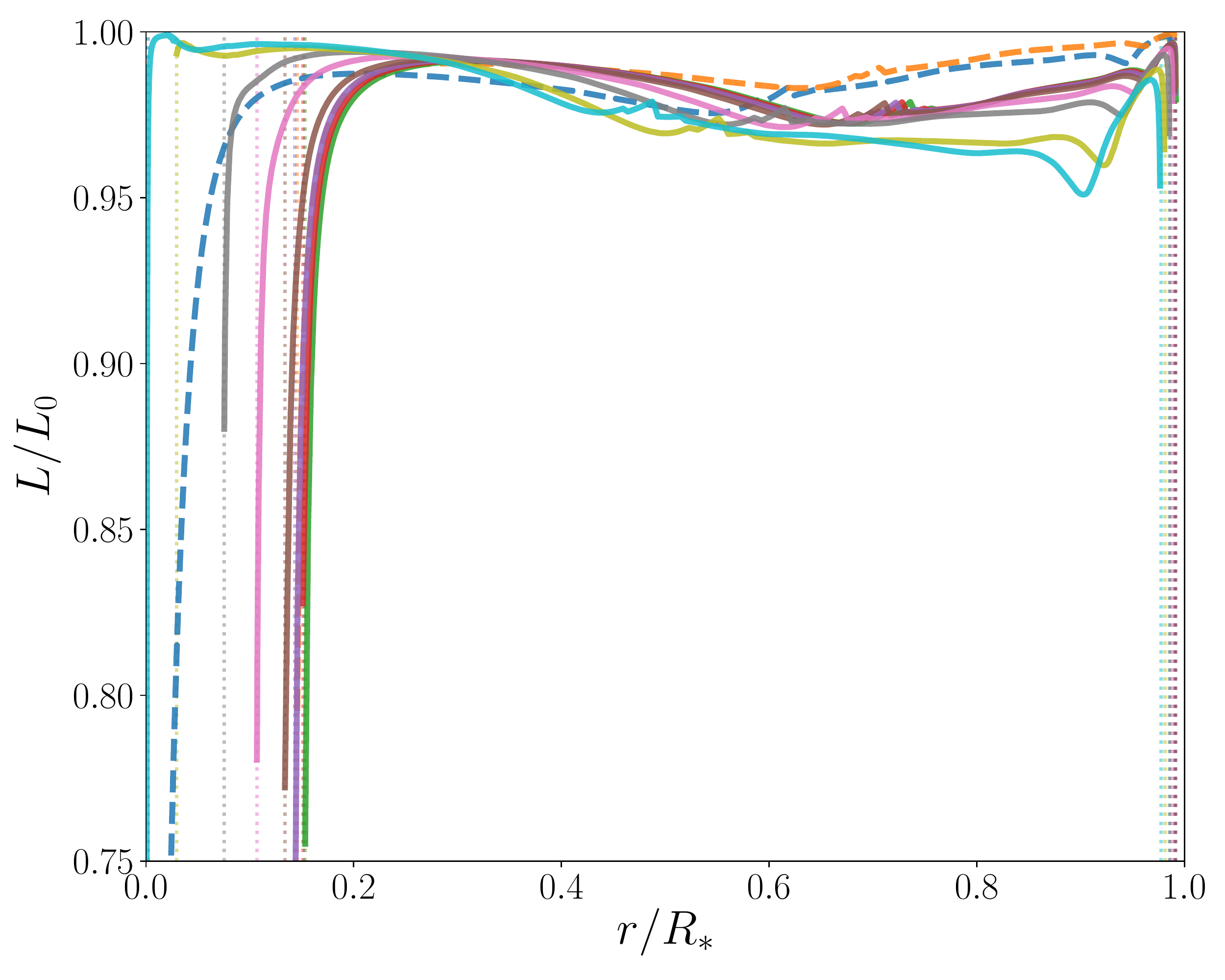}\hspace{-0.15cm}
    \includegraphics[height=6.35cm]{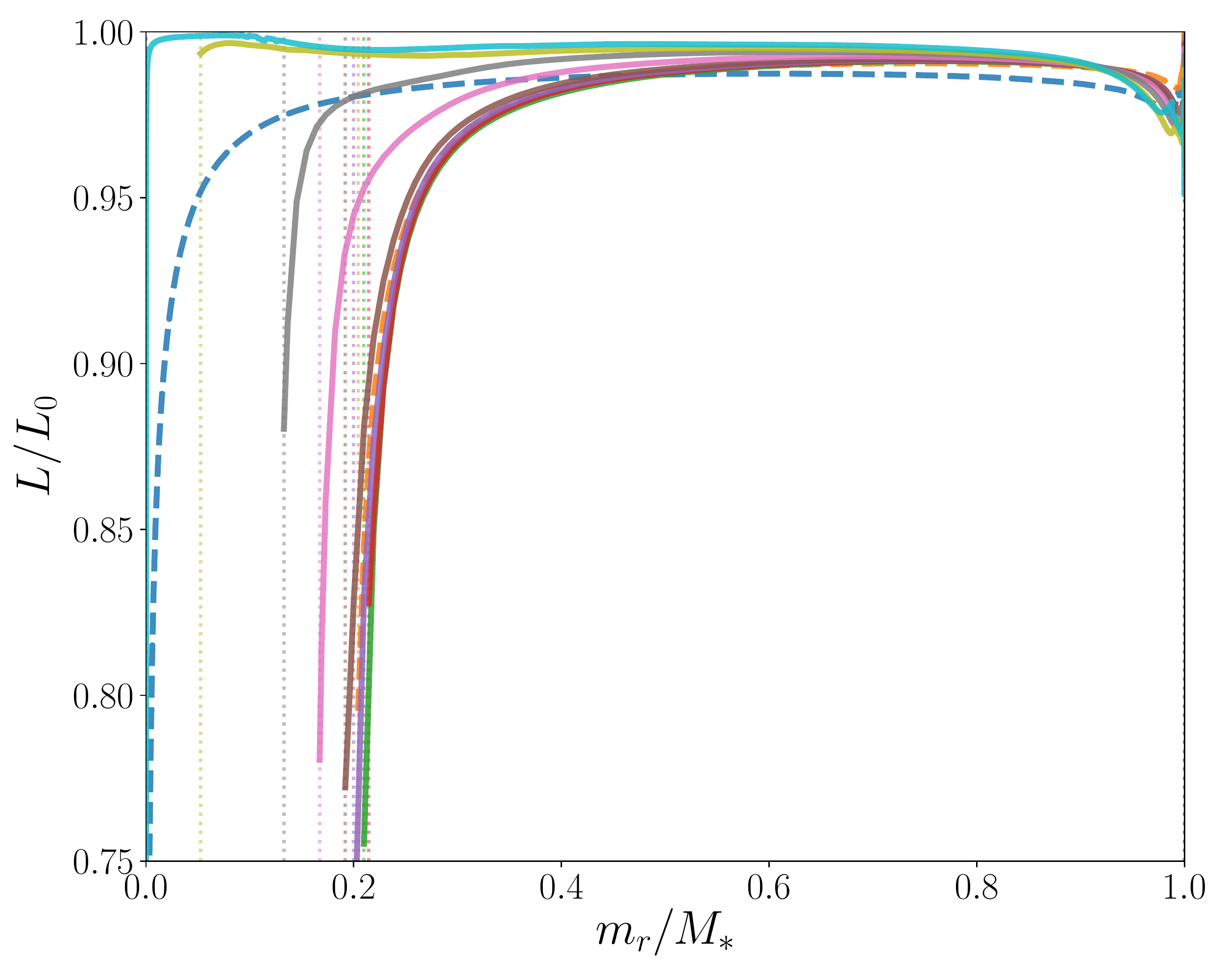}\hspace{-0.2cm}
    \begin{tikzpicture}[scale=1.0]
     \node[color=white] at (0,0) {o};
     \node[anchor=south west,inner sep=0] at (0,1.4) {\includegraphics[height=3.5cm]{colorbar.png}};
    \end{tikzpicture}\\[-0.3cm]
    \caption{Same as Fig.~\ref{fig:S} but for the wave penetration length, $L$, normalised by that of pure gravity waves, $L_0$. This was calculated from Eq.~(\ref{eq:L}).}
    \label{fig:L}
    %
\end{figure*}
\begin{figure*}[ht]
    \centering
    \includegraphics[height=6.35cm]{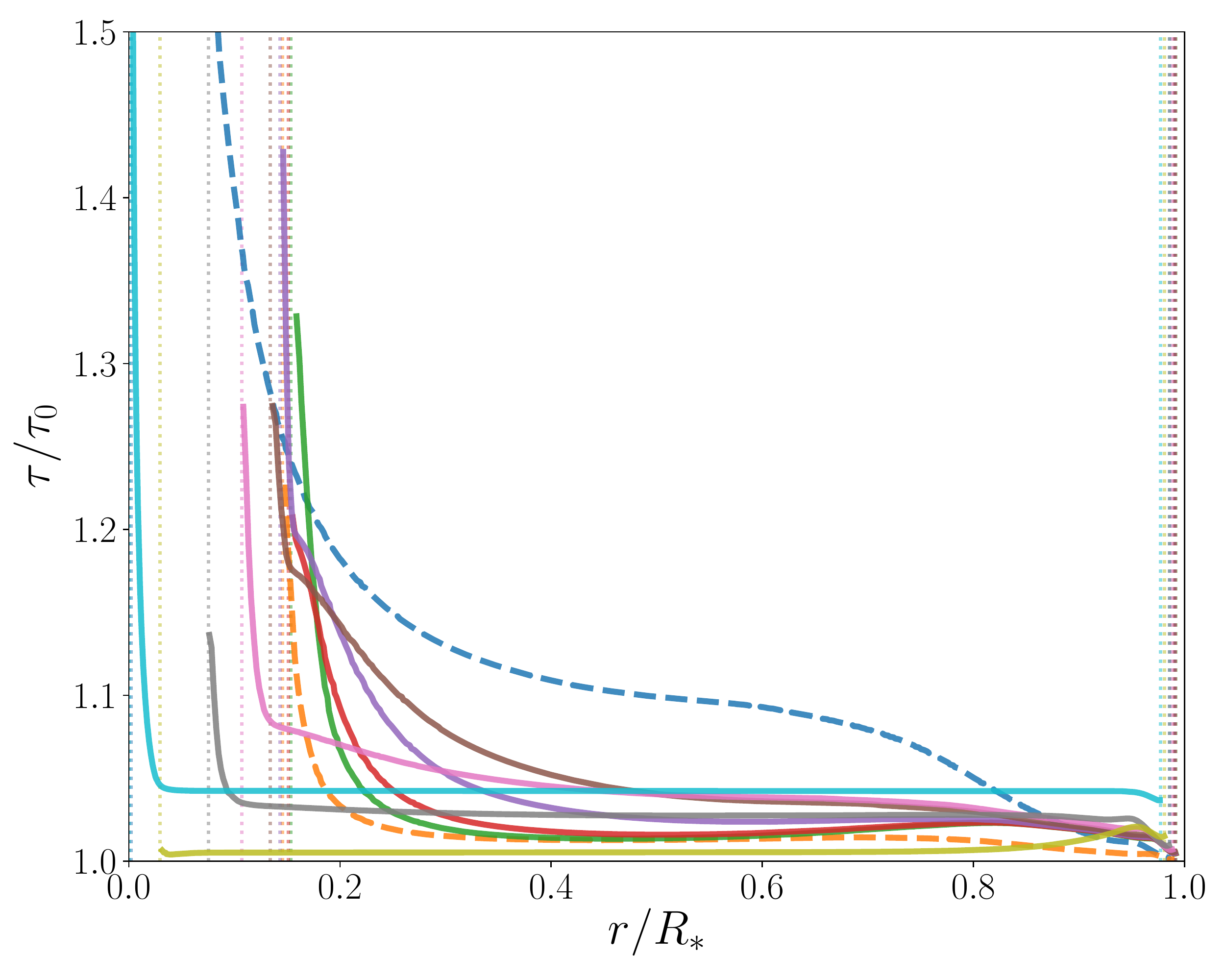}\hspace{-0.15cm}
    \includegraphics[height=6.35cm]{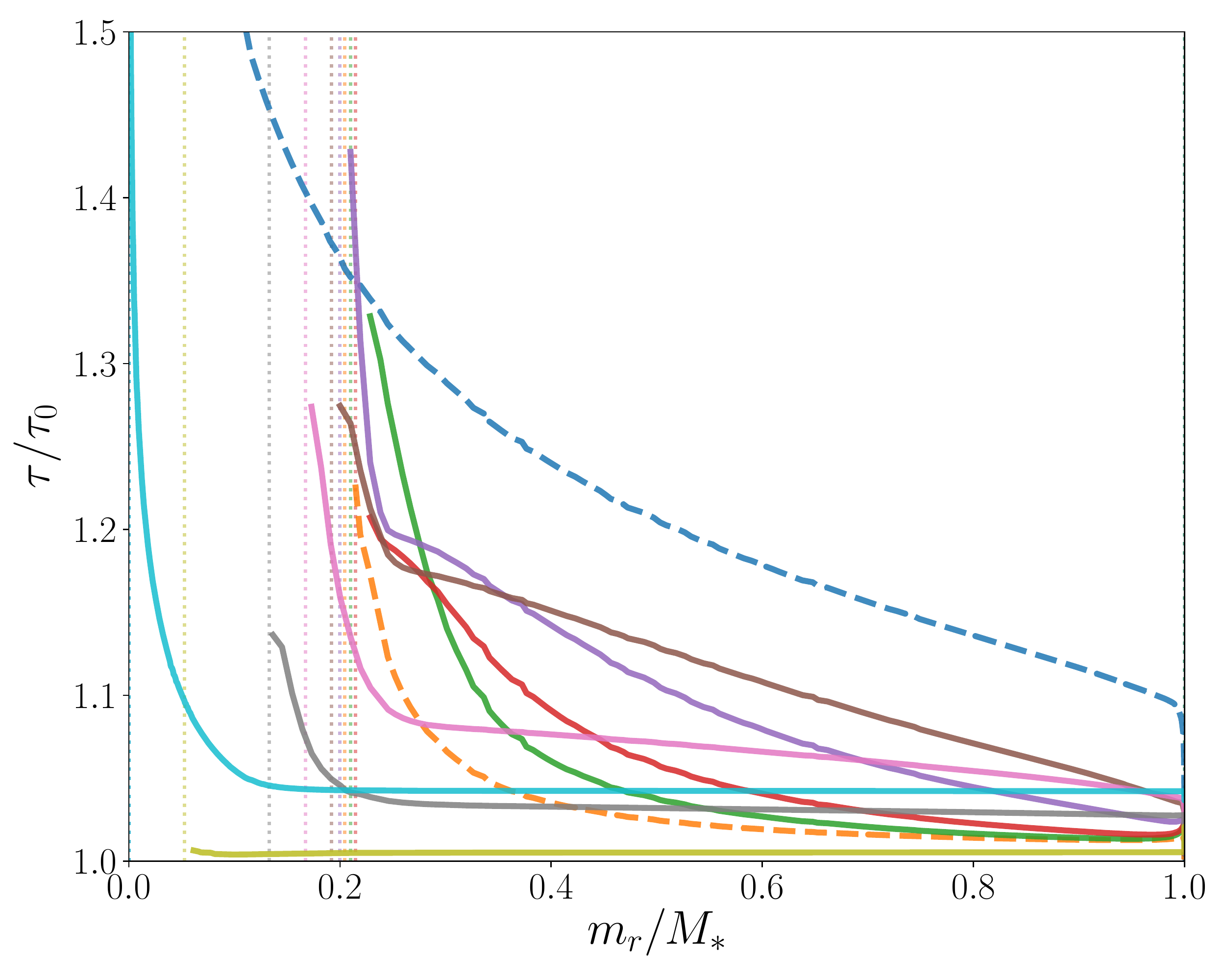}\hspace{-0.2cm}
    \begin{tikzpicture}[scale=1.0]
     \node[color=white] at (0,0) {o};
     \node[anchor=south west,inner sep=0] at (0,1.4) {\includegraphics[height=3.5cm]{colorbar.png}};
    \end{tikzpicture}\\[-0.3cm]
    \caption{Same as Fig.~\ref{fig:S} but for the damping, $\tau$, normalised by that of pure gravity waves, $\tau_0$. This was calculated from Eq.~(\ref{eq:tau}).}
    \label{fig:tau}
\end{figure*}

First, we show how our prescription for the penetration length, given by Eq.~(\ref{eq:L}), depends on radius and age. For that purpose, Fig.~\ref{fig:L} shows the profiles of $L/L_0$, calculated from Eq.~(\ref{eq:L}), for our different models presented on Figs. \ref{fig:S} \& \ref{fig:Rim12}. 
These have been calculated with $\textrm{Fr=0.01}$, $l=5$ and $m=1$ (corresponding to $\alpha=24/25$, see Eq.~(\ref{eq:alpha})).
It can be seen \citep[like in the solar-case of][]{AndreEtal2018_pnps}, that the penetration length $L$ (with rotation) is always lower than the one for pure gravity waves in our parameters range, through the whole radiative region. This means that thermal damping is found to be more efficient when rotation effects are included in the analysis. Moreover, except very close to the convective core, there is less than a 5\% difference between both prescriptions in the bulk of the radiative zone.
However, the penetration length of gravito-inertial waves is significantly decreased by rotation in the very inner part of the radiative zone, near the excitation region. This is where the buoyancy frequency {sharply decreases and} smoothly matches its small negative value of the nearby turbulent convection core.
The penetration length when including rotation reaches a 10\% to 30\% discrepancy at the base of the radiative zone, through the whole MS. 

The analysis of the wave penetration length gives us a local understanding of how thermal diffusion acting on internal gravity waves differs when rotation effects are included, under our assumptions. In order to examine the integrated effect along their propagation, we now turn to compute the damping factor of gravito-inertial waves, $\tau$, normalised by that of pure gravity waves, $\tau_0$. These are computed from Eqs.~(\ref{eq:tau}) \& (\ref{eq:L}). The resulting profiles are shown on Fig.~\ref{fig:tau}, which have been obtained with $\textrm{Fr}=(1\,\mu\textrm{Hz})/N$, $l=5$ and $m=1$ (corresponding to $\alpha=24/25$, see Eq.~(\ref{eq:alpha})). As a first step, we have neglected the Doppler-shift of the wave's frequency. As expected from the behaviour of the wave penetration length, the damping factor is greater when rotation effects are included, which leads to $\tau>\tau_0$ for all our models. We also note that $\tau/\tau_0$ is globally a decreasing function of radius: the discrepancy between $\tau$ and $\tau_0$ is higher in the inner radiative region, and narrows down to close to unity near the surface. Thus, the amplitude of the wave when arriving at the surface is not significantly changed, but the redistribution of AM could be altered in inner regions. {This is of important interest since these layers close to the core boundary are those which are probed by asteroseismology \citep[e.g.][]{VanReethEtal2016,VanReethEtal2018}.}

{Coming back to the mathematical analysis detailed in \cite{AndreEtal2018_pnps}, we can give some insight on why the effect of rotation is weak on average. Focusing on Eq.~(2.5) of \cite{AndreEtal2018_pnps}, we see that thermal diffusion acts on the ratio $\rho'/\rho$ (where $\rho'$ is the density fluctuation and $\rho$ is the density background) which is the driving term of the buoyancy force \cite[in the radial momentum equation, see Eq.~(2.1) of][]{AndreEtal2018_pnps}. However, thermal diffusion does not act directly on the Coriolis acceleration in a similar way. Because the buoyancy force (characterised by the buoyancy frequency $N$) dominates over the Coriolis acceleration (characterised by the Coriolis frequency $2\Omega$) in the radial direction, we can understand why rotation does not have a great impact upon the resulting thermal damping of waves, calculated from the linearised set of primitive equations.}

\section{Discussion \& conclusions}\label{sec:conclusion}
 Using the analysis presented in \cite{AndreEtal2018_pnps}, in an equatorial model, we found that the penetration length of gravito-inertial waves associated to their thermal diffusion is not significantly modified by rotation on average in the case of a $3\,M_{\odot}$ intermediate-mass star. The resulting integrated damping of a wave launched at the core boundary has not found to be strongly increased at the stellar surface. In the framework of this analysis, we expect this statement to be valid for any intermediate-mass star, as our control parameters should lie in the same range of values. Consequently, angular momentum exchanges through thermal diffusion is not a physical process that is expected to be strongly modified by global and differential rotation. 
 
 However, we have shown that the wave penetration length can be strongly decreased by rotation at the boundary between the convective core and the radiative envelope. This is where waves are excited by convective penetration or overshoot. Therefore, stochastically excited gravito-inertial waves are expected from our analysis to be damped more strongly near their excitation region, compared to pure gravity waves. This confirms the same result obtained by \cite{PantillonEtal2007} using the TAR. 
 
 Finally, we recall that one should also evaluate the effects of (differential) rotation on corotation resonances \cite[e.g.][]{AlvanEtal2013}, wave nonlinear breaking \citep[e.g.][]{Press1981,BarkerOgilvie2010,RatnasingamEtal2019}, and wave stochastic excitation \citep[e.g.][]{MathisEtal2014}, to be able to draw a coherent and quantitative theoretical picture and comparison with direct 2D equatorial numerical simulations. Moreover, as a next step, 3D complex behavior such as latitudinal trapping that modify the energy transmission from convection to waves and their propagation \citep[e.g.][]{Mathis2009,Prat2018} should be taken into account.

\section*{Acknowledgments}
QA and SM acknowledge funding by the European Research Council through ERC SPIRE grant 647383, and the PLATO CNES grant at CEA-Saclay. LA also thanks the ERC through grant 682393 (AWESoMeStars).

\bibliographystyle{phostproc}
\bibliography{stars.bib}

\end{document}